\documentclass[preprint,12pt]{elsarticle}

\usepackage{amssymb}

\usepackage{amsmath,amsthm,amssymb,mathrsfs}
\usepackage{bm}

\journal{Journal of Magnetism and Magnetic Materials}

\begin{document}

\begin{frontmatter}




\title{Phase locking in voltage-controlled parametric oscillator}


\author{Tomohiro Taniguchi 
}


\address{
 National Institute of Advanced Industrial Science and Technology (AIST), Research Center for Emerging Computing Technologies, Tsukuba, Ibaraki 305-8568, Japan, 
}



\begin{abstract}
A recent experimental demonstration of a parametric magnetization oscillation excited by applying a microwave voltage to a ferromagnetic metal will be applicable not only to a new magnetization switching method but also to bio-inspired computing. 
It should be, however, noted that a phase of the parametric magnetization oscillation is not uniquely locked, related to the fact that a frequency of the microwave voltage is twice the value of the magnetization oscillation. 
There are two possible phases in the parametric oscillation state, and which of the two is realized depends on the initial condition of the magnetization. 
Here, we examine two approaches to lock the phase uniquely. 
One is to suppress the distribution of the initial state by enhancing the perpendicular magnetic anisotropy before applying microwave voltage, and the other is to use a sweeping frequency. 
Through numerical simulation of the Landau-Lifshitz-Gilbert equation and quantification of locked rate, we find that the sweeping frequency is more effective to lock the phase of the parametric magnetization oscillation. 
\end{abstract}


\end{frontmatter}





\section{Introduction}
\label{sec:Introduction}

Magnetization dynamics studied in magnetism and spintronics are mainly classified into two groups, magnetization switching and oscillation. 
The magnetization switching is excited by applying magnetic field \cite{okamoto12,taniguchi14}, electric current \cite{slonczewski96,berger96,katine00,myers02,krivorotov05,taniguchi16}, and/or voltage \cite{maruyama09,shiota09,nozaki10,endo10,shiota11,wang11,shiota12,kanai12,shiota13,grezes16,shiota17,nozaki18,okada18} to a ferromagnet, and the techinique of which has been applied to non-volatile memory applications \cite{dieny16}. 
The magnetization oscillation is, on the other hand, driven by a microwave magnetic field \cite{vonsovskii66} or by applying direct or oscillating electric current \cite{kiselev03,rippard04,tulapurkar05,houssameddine07,kubota08,sankey08,kubota13}, and is expected to be used in micro- and mill-wave sensors and generators \cite{zhou19,kurokawa22} and bio-inspired computing \cite{grollier16,torrejon17,kudo17,yamaguchi20,grollier20,prasad22,yamaguchi23}. 
Note that the magnetization oscillation requires a continuous energy injection into a ferromagnet to sustain the oscillation against energy dissipation due to damping torque. 
It has been difficult to excite an oscillation by an application of voltage because voltage controlled magnetic anisotropy (VCMA) effect merely changes the shape of magnetic potential energy and does not act as an energy injector. 
Recently, however, an experimental demonstration of a parametric magnetization oscillation through the VCMA effect was reported \cite{yamamoto20}, where the magnetization oscillated with Larmor frequency $f_{\rm L}$ when microwave voltage with a frequency of $2f_{\rm L}$ was applied. 
Such method might solve issues in oscillator devices driven by electric current, such as large energy dissipation due to Joule heating. 


In the parametric magnetization oscillation by the VCMA effect, on the other hand, since the frequency of the driving voltage is twice the value of the magnetization oscillation, two possible phases exist in a steady state. 
Accordingly, the oscillating output from the ferromagnet is not unique \cite{taniguchi22}. 
The situation should be avoided for some practical applications, such as, physical reservoir computing \cite{torrejon17,yamaguchi20}, where one-to-one correspondence between input and output signal, called echo state property \cite{yildiz12}, is required \cite{yamaguchi23,imai22}. 
This is in contrast with a phase locking of electric-current-driven oscillators \cite{quinsat11,rippard13}, where the frequency of the magnetization oscillation becomes identical to that of input signal and the phase with respect to the input signal is uniquely locked. 


In this work, we performed numerical and theoretical analyses for a phase locking of a parametric oscillator driven by the VCMA effect. 
First, we show that the phase of the oscillation depends on the initial state of the magnetization, which usually fluctuates due to thermal activation. 
The distribution of the initial state can be suppressed by enhancing a perpendicular magnetic anisotropy before the application of microwave voltage. 
This approach is, however, not effective because a tiny difference in the initial state leads to a different phase. 
The other proposal is to sweep the frequency of the microwave voltage. 
When the frequency of the voltage is slightly different from $2f_{\rm L}$, the relative phase between the magnetization oscillation and the voltage is uniquely locked because of an asymmetry in the stability of the phase. 
The phase is locked even after changing the frequency of the voltage to $2f_{\rm L}$. 
These results are obtained by solving the Landau-Lifshitz-Gilbert (LLG) equation both numerically and analytically. 



\begin{figure}
\centerline{\includegraphics[width=1.0\columnwidth]{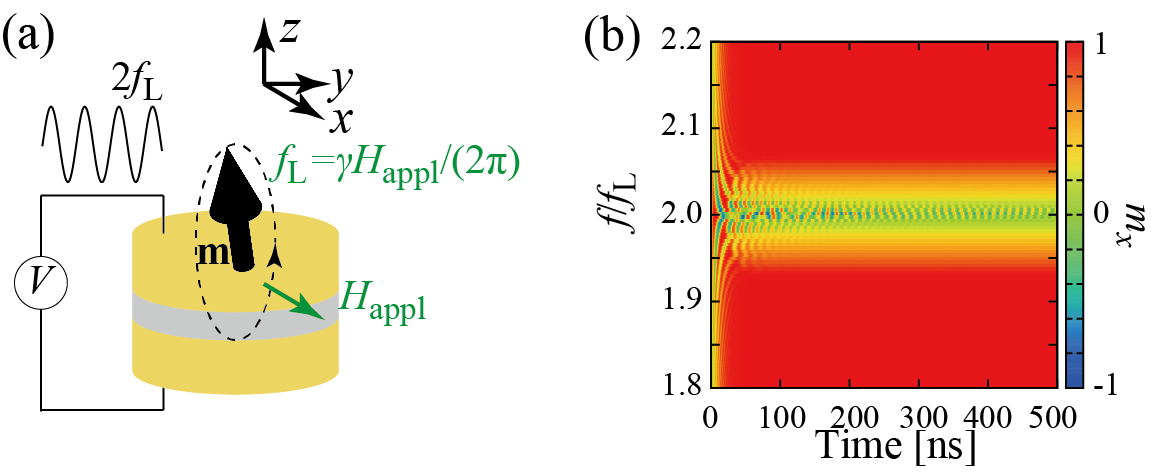}}
\caption{
            (a) Schematic illustration of a parametric oscillation of magnetization in a ferromagnetic multilayer. 
                 A unit vector pointing in the magnetization direction in free layer is $\mathbf{m}$. 
                 The $z$ axis is normal to the film plane, while the $x$ axis is parallel to an external magnetic field. 
                 It oscillates around the external magnetic field $H_{\rm appl}$ with the Larmor frequency $f_{\rm L}=\gamma H_{\rm appl}/(2\pi)$ when microwave voltage with a frequency of $2f_{\rm L}$ is applied. 
           (b) Time evolution of $m_{x}$ in the presence of the microwave voltage. 
                The vertical axis represents a ratio of the frequency $f$ of the microwave voltage to the Larmor frequency. 
         \vspace{-3ex}}
\label{fig:fig1}
\end{figure}


\section{System description}
\label{sec:System description}


\subsection{LLG equation}
\label{sec:LLG equation}

\begin{figure}
\centerline{\includegraphics[width=1.0\columnwidth]{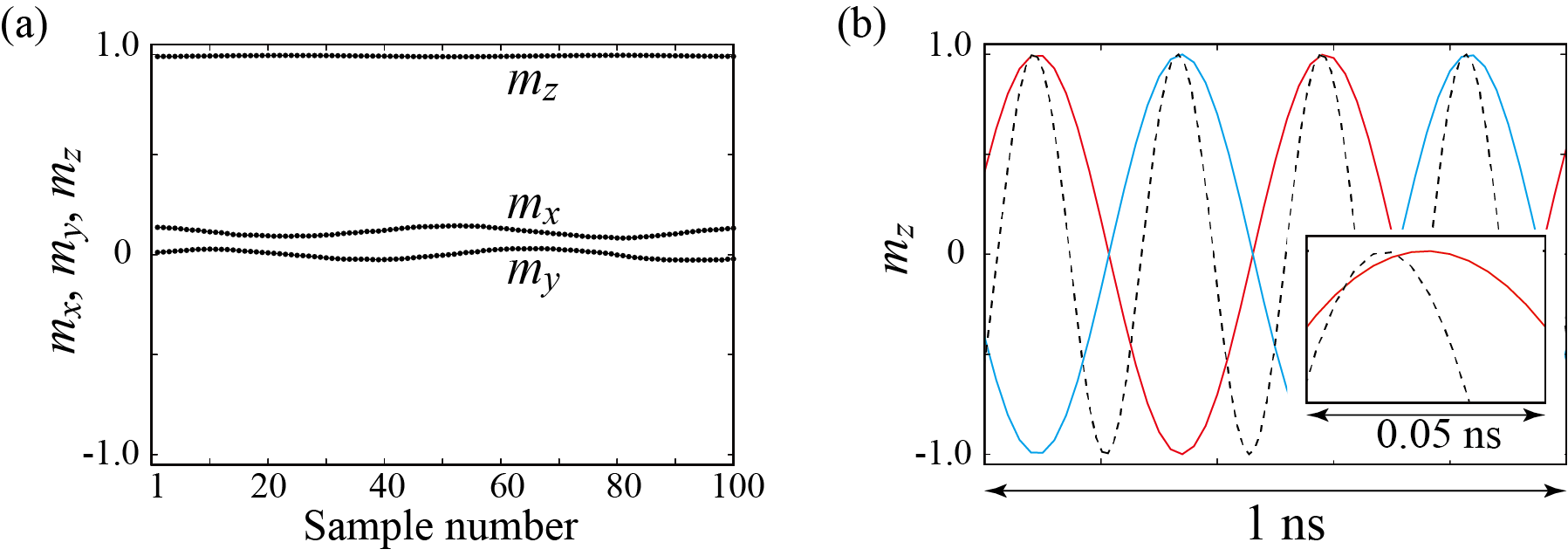}}
\caption{
            (a) $100$ samples of an initial state prepared by solving the LLG equation with thermal activation. 
            (b) Examples of oscillation of $m_{z}$ for two different initial states. 
                The dotted line represents the oscillation of the microwave voltage. 
                The blown up inset indicates that the peak positions of $m_{z}$ and the microwave voltage are slightly different. 
         \vspace{-3ex}}
\label{fig:fig2}
\end{figure}

In Fig. \ref{fig:fig1}(a), we show a schematic illustration of a ferromagnetic/nonmagnetic/ferromagnetic trilayer. 
The top and bottom ferromagnets are free and reference layers, respectively. 
The unit vector pointing in the magnetization direction in the free layer is denoted as $\mathbf{m}$, where it has been experimentally confirmed that a macrospin model works well to describe the magnetization dynamics due to the VCMA effect \cite{yamamoto20}. 
The thickness of the nonmagnetic insulating layer is thick enough so that charge accumulation is generated near the interface when an electric voltage is applied. 
The charge accumulation modulates electron-states and change the magnetic anisotropy \cite{duan08,nakamura09,tsujikawa09}. 
Accordingly, an application of voltage changes a stable state of the magnetization and drives the magnetization dynamics, which is described by the LLG equation, 
\begin{equation}
  \frac{d \mathbf{m}}{dt}
  =
  -\gamma
  \mathbf{m}
  \times
  \mathbf{H}
  +
  \alpha 
  \mathbf{m}
  \times
  \frac{d \mathbf{m}}{dt}, 
  \label{eq:LLG}
\end{equation}
where $\gamma$ and $\alpha$ are the gyromagnetic ratio and the Gilbert damping constant, respectively. 
For the parametric oscillation induced by the VCMA effect, the magnetic field $\mathbf{H}$ is given by 
\begin{equation}
  \mathbf{H}
  =
  H_{\rm appl}
  \mathbf{e}_{x}
  +
  H_{\rm K}
  m_{z}
  \mathbf{e}_{z}, 
  \label{eq:field}
\end{equation}
with 
\begin{equation}
  H_{\rm K}
  =
  H_{\rm Ka}
  \sin\left(2\pi ft \right). 
  \label{eq:HK_microwave}
\end{equation}
Here, $H_{\rm appl}$ is an external magnetic field applied in the $x$ direction, while $H_{\rm K}$ is the perpendicular magnetic anisotropy field along the $z$ axis. 
In Eq. (\ref{eq:HK_microwave}), $H_{\rm K}$ has only an oscillating component $H_{\rm Ka}$ with a frequency of $f$, and a direct component $H_{\rm Kd}$ is assumed to be zero, for simplicity. 
Such a situation can be experimentally realized by applying both direct and microwave voltages to the free layer \cite{yamamoto20}. 
In Fig. \ref{fig:fig1}(b), we show time evolution of $m_{x}$ obtained by solving Eq. (\ref{eq:LLG}) numerically, where the values of the parameters are derived from typical experiments as $\gamma=1.764 \times 10^{7}$ rad/(Oe s), $\alpha=0.005$, $H_{\rm appl}=720$ Oe, and $H_{\rm Ka}=100$ Oe (see also \ref{sec:AppendixA} for the details of the numerical simulations). 
It is shown that $m_{x}$ tends to be zero when $f$ is close to twice the value of the Larmor frequency $f_{\rm L}=\gamma H_{\rm appl}/(2\pi)$. 
In this case, the magnetization oscillates around the $x$ axis (see also Fig. \ref{fig:fig2} discussed below).  
On the other hand, when $f$ differs from $2 f_{\rm L}$, $m_{x}$ saturates to $+1$, which indicates that the magnetization relaxes to the direction of the external magnetic field. 


\subsection{Dependence of phase on initial state}
\label{sec:Dependence of phase on initial state}

Next, we show that the phase of the parametric oscillation is not uniquely locked. 
Before applying microwave voltage, the magnetic field is given by $\mathbf{H}=H_{\rm appl}\mathbf{e}_{x}+H_{\rm Kd}m_{z}\mathbf{e}_{z}$, where the direct component $H_{\rm Kd}$ of the perpendicular magnetic anisotropy field consists of shape and bulk-and-interfacial magnetic anisotropy fields. 
The magnetization points to a direction close to an energetically stable state, $\mathbf{m}^{(0)}=[m_{x}^{(0)},m_{y}^{(0)},m_{z}^{(0)}]=[(H_{\rm appl}/H_{\rm Kd}),0,\pm\sqrt{1-(H_{\rm appl}/H_{\rm Kd})^{2}}]$ with an assumption $H_{\rm appl}/H_{\rm Kd}<1$, at which an energy density $E=-M\int d \mathbf{m}\cdot\mathbf{H}$ is minimized ($M$: saturation magnetization). 
The magnetization shows a small-amplitude oscillation around this equilibrium direction $\mathbf{m}^{(0)}$ due to thermal activation. 
Therefore, when we apply the voltage, the initial state of the magnetization is randomly distributed. 
To estimate such a distributed initial condition, we solve the LLG equation with $\mathbf{H}=H_{\rm appl}\mathbf{e}_{x}+H_{\rm Kd}m_{z}\mathbf{e}_{z}$ and a random torque $-\gamma \mathbf{m}\times\mathbf{h}$. 
The components $h_{k}$ ($k=x,y,z$) of the random field $\mathbf{h}$ satisfy the fluctuation-dissipation theorem \cite{brown63}, 
\begin{equation}
  \langle h_{k}(t) h_{\ell}(t^{\prime}) \rangle 
  =
  \frac{2\alpha k_{\rm B}T}{\gamma MV}
  \delta_{k\ell}
  \delta(t-t^{\prime}), 
\end{equation}
where $V=Sd$ is the volume of the free layer consisting of a cross-section area $S$ and thickness $d$. 
In this work, we use $M=955$ emu/cm${}^{3}$, $H_{\rm Kd}=6.283$ kOe, and $d=1.1$ nm from Ref. \cite{yamamoto20}, while $S$ is assumed to be $\pi\times 50^{2}$ nm${}^{2}$ \cite{yamamoto20}. 
Temperature $T$ is $300$ K. 
Note that the random torque is included in the LLG equation only when we evaluate the initial state, whereas it is neglected during the calculation of the parametric oscillation to clarify the roles of our suggestions developed in Secs. \ref{sec:Suppression of initial distribution} and \ref{sec:Results of numerical simulation}. 
The role of thermal activation during the parametric oscillation state will be studied in Sec. \ref{sec:Role of thermal activation}.

The calculated results of $100$ samples of the initial state obtained by this method is shown in Fig. \ref{fig:fig2}(a). 
Since $H_{\rm Kd}\gg H_{\rm appl}$, the magnetization points approximately to the $z$ direction with slight shift to the $x$ direction. 
Next, let us go back to the parametric oscillation state, where the initial condition of Eq. (\ref{eq:LLG}) is chosen from one of  these $100$ samples and the frequency of the microwave voltage is $2f_{\rm L}$. 
Two solid lines in Fig. \ref{fig:fig2}(b) show examples of the parametric oscillation of $m_{z}$ with two different initial conditions. 
We also show the oscillation of the microwave voltage by a dotted line. 
As can be seen, the magnetization oscillates with half the frequency of the microwave voltage, i.e., $f_{\rm L}$, and there are two possible phases of the magnetization with respect to the microwave voltage, depending on the initial state. 
Since the initial state is usually uncontrollable, the result indicates that the phase in the parametric oscillation is not uniquely locked. 
To solve the issue, we examine two approaches. 
The first one is to suppress the distribution of the initial state, and the other is to use a sweeping frequency. 
In the following sections, we describe the details of these approaches and their effectiveness. 


\section{Suppression of initial distribution}
\label{sec:Suppression of initial distribution}

The results shown in Figs. \ref{fig:fig2}(a) and \ref{fig:fig2}(b) indicate that the randomness of the initial state prevents from locking the phase uniquely. 
We notice that the distribution of the initial state can be suppressed by the VCMA effect due to the following reason. 
It has been experimentally revealed \cite{maruyama09,shiota09,nozaki10,endo10,shiota11,wang11,shiota12,kanai12,shiota13,grezes16,shiota17,nozaki18,okada18} that the perpendicular magnetic anisotropy is reduced when an electric field generated by applying a voltage points to a one direction, while it is enhanced when the field points to the opposite direction. 
In other words, the perpendicular magnetic anisotropy can be either reduced or enhanced, depending on the sign of the voltage. 
In the switching and parametric-oscillation applications, the sign of the direct voltage is chosen so that the perpendicular magnetic field is reduced in order to move the magnetization from an initial state and starting its dynamics. 
On the other hand, when a voltage with an opposite sign is applied, the magnetization moves to the direction of the nearest equilibrium state, i.e., $\mathbf{m}$ moves to the direction of $\mathbf{m}^{(0)}$ against thermal activation. 
In this case, the distribution of the initial state will be suppressed. 
This method was used to reduce a switching error for memory applications \cite{yamamoto20PRA}, where, before applying a voltage for the switching, another voltage having an opposite sign is applied.  


\begin{figure}
\centerline{\includegraphics[width=1.0\columnwidth]{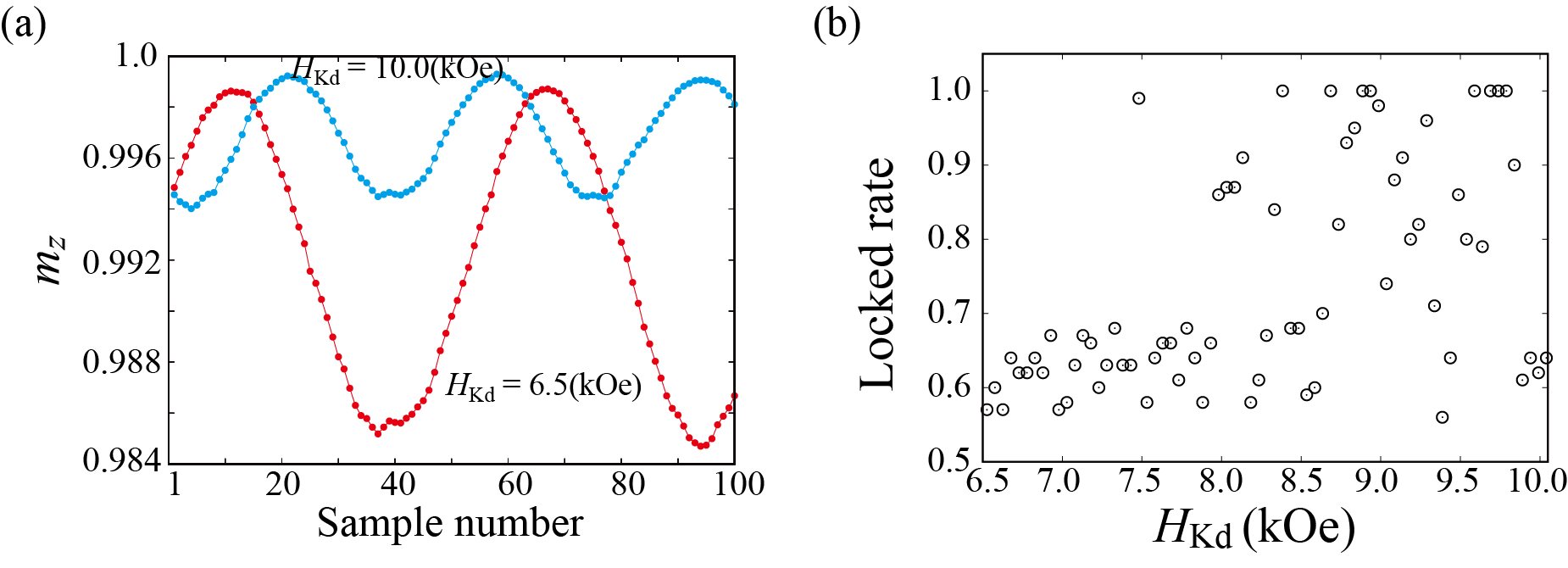}}
\caption{
           (a) The $z$ components of initial conditions of $100$ samples, where $H_{\rm Kd}\simeq 6.5$ ($10.0$) kOe for red (blue) dots.  
           (b) Locked rate as a function of $H_{\rm Kd}$. 
         \vspace{-3ex}}
\label{fig:fig3}
\end{figure}


Here, we apply the method shown above to suppress the distribution of the initial state. 
As mentioned, $100$ samples of the initial state are prepared by solving the LLG equation with the magnetic field of $\mathbf{H}=H_{\rm appl}\mathbf{e}_{x}+H_{\rm Kd}m_{z}\mathbf{e}_{z}$. 
We assume that $H_{\rm Kd}$ of which before applying the microwave voltage, is enhanced due to the VCMA effect caused by a direct voltage with an appropriate sign. 
In Fig. \ref{fig:fig3}(a), we show two examples of the initial $m_{z}$ obtained for $H_{\rm Kd}\simeq 6.5$ (red) and $10.0$ kOe (blue). 
It is shown that the distribution of the initial state is suppressed for a large $H_{\rm Kd}$. 
Using these initial conditions, we performed the numerical simulation of the LLG equation and investigated the phase in the parametric oscillation state. 

To quantify the uniqueness of the phase, we introduce a locked rate ${\rm LR}$ as follows; 
\begin{equation}
  {\rm LR}
  =
  \frac{{\rm max}[N_{+},N_{-}]}{N_{+}+N_{-}}, 
\end{equation}
where $N_{+}$ is the number of the samples having a same phase, while $N_{-}$ is the number of the samples having another same phase. 
Since the phase in the parametric oscillation state is locked to either one of them, $N_{+}+N_{-}$ equals to the sample numbers ($100$ in this study). 
When the phase of the magnetization becomes independent of the initial state and is uniquely locked, ${\rm LR}=1$, while ${\rm LR}$ becomes $0.5$ when two possible phases are equally realized. 
In Fig. \ref{fig:fig3}(b), we summarize the dependence of the locked rate ${\rm LR}$ on the perpendicular magnetic anisotropy field $H_{\rm Kd}$ before applying the microwave voltage. 
Roughly speaking, the locked rate increases as $H_{\rm Kd}$ increases. 
However, the locked rate is widely distributed even for relatively large $H_{\rm Kd}$. 
The result indicates that the phase in the parametric oscillation state is sensitive to the initial state, even after its distribution is suppressed. 
Therefore, we conclude that this method is not effective enough for phase locking. 


\section{Phase locking by using sweeping frequency}
\label{sec:Phase locking by using sweeping frequency}

\subsection{Theoretical aspect of proposal}
\label{sec:Theoretical aspect of proposal}

In this section, we attempt another method to lock the phase uniquely by using a sweeping frequency, where the frequency of the microwave voltage initially is slightly different from $2f_{\rm L}$ and slowly changes to $2f_{\rm L}$. 
The point of this method is as follows. 
The fact that there are two possible phases in a steady state implies that there is a double-well of the phase and the depths of two minima are equal. 
he situation can be confirmed by deriving an approximated equation of motion for the phase from Eq. (\ref{eq:LLG}). 
We introduce a spherical coordinate $(\Theta,\Phi)$ from $\mathbf{m}$ as $\mathbf{m}=(m_{x},m_{y},m_{z})=(\cos\Theta,\sin\Theta\cos\Phi,\sin\Theta\sin\Phi)$. 
In terms of $\Theta$ and $\Phi$, the LLG equation is explicitly given by 
\begin{equation}
\begin{split}
  \frac{d\Theta}{dt}
  =&
  \gamma
  H_{\rm Ka}
  \sin(2\pi ft)
  \sin\Theta
  \sin\Phi
  \cos\Phi
\\
  &-
  \alpha
  \gamma
  \left[
    H_{\rm appl}
    -
    H_{\rm Ka}
    \sin(2\pi ft)
    \cos\Theta
    \sin^{2}\Phi
  \right]
  \sin\Theta,
  \label{eq:LLG_theta}
\end{split}
\end{equation}
\begin{equation}
\begin{split}
  \frac{d\Phi}{dt}
  =&
  \gamma
  H_{\rm appl}
  -
  \gamma
  H_{\rm Ka}
  \sin(2\pi ft)
  \cos\Theta
  \sin^{2}\Phi
\\
  &+
  \alpha
  \gamma 
  H_{\rm Ka}
  \sin(2\pi ft)
  \sin\Phi
  \cos\Phi, 
  \label{eq:LLG_phi}
\end{split}
\end{equation}
where we use an approximation $1+\alpha^{2}\simeq 1$, for simplicity. 
Since we are interested in dynamics in which the magnetization oscillates with a frequency nearly half of that of the microwave voltage $f$, we introduce $\Psi$ as 
\begin{equation}
  \Psi
  =
  \Phi
  -
  \pi ft . 
  \label{eq:definition_psi}
\end{equation}
Note that the phase of the microwave voltage is $2\pi ft$, while $\pi ft$ appears in Eq. (\ref{eq:definition_psi}). 
While $\Phi-2\pi ft$ is not a constant, $\Psi=\Phi-\pi ft$ becomes an approximately constant when the microwave frequency $f$ is close to $2 f_{\rm L}$. 
Let us investigate this point in the following. 

\begin{figure}
\centerline{\includegraphics[width=1.0\columnwidth]{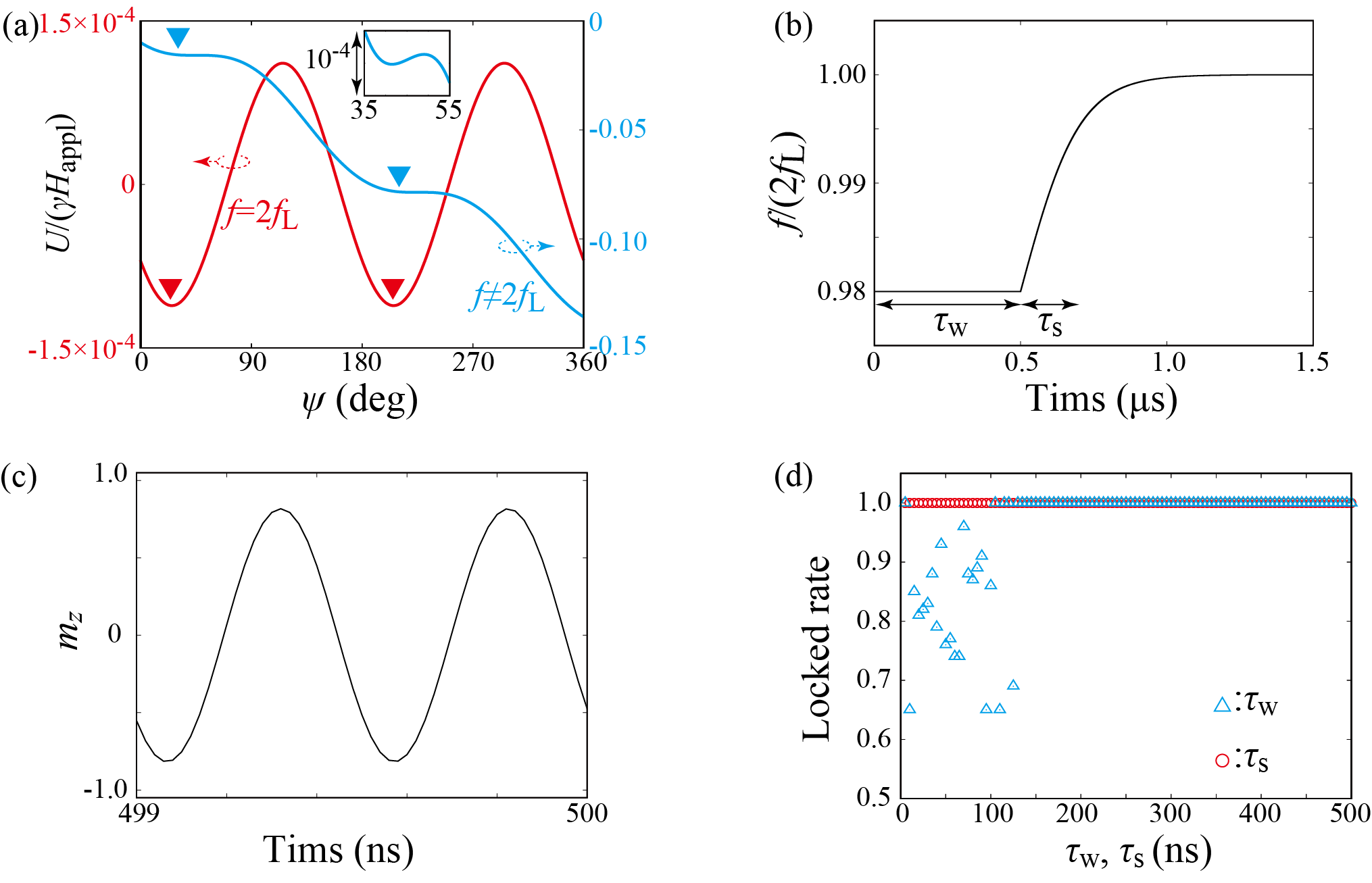}}
\caption{
           (a) Schematic illustration of potential maps for $f=2f_{\rm L}$ (red) and $f\neq 2f_{\rm L}$ ($f=2f_{\rm L}\times 0.98 $) (blue). 
                The potential $U$ is normalized by $\gamma H_{\rm appl}$. 
                The averaged $\cos\Theta$ is estimated from the numerical simulation of the LLG equation as $\cos\Theta\simeq 0.004$ for $f=2f_{\rm L}$ and $0.581$ for $f=2f_{\rm L}\times 0.98$. 
                Triangles indicate the positions of local minima. 
                The inset is an enlarged view of the potential for $f=2f_{\rm L}\times 0.98$ near its local minimum. 
           (b) An example of time evolution of the frequency of microwave-voltage, $f/(2f_{\rm L})$, where $\tau_{\rm w}=500$ ns and $\tau_{\rm s}=200$ ns. 
                The frequency is slightly different from $2f_{\rm L}$ when time $t$ is less than waiting time $\tau_{\rm w}$, and saturates to $2f_{\rm L}$ with a time scale of sweeping time $\tau_{\rm s}$. 
           (c) Time evolution of $m_{z}$ when the frequency is $2rf_{\rm L}$. 
           (d) Locked rate as a function of $\tau_{\rm w}$ and $\tau_{\rm s}$. 
         \vspace{-3ex}}
\label{fig:fig4}
\end{figure}

Using an approximation to average the equation of motion with respect to a fast variable, i.e., averaging the equation with respect to time over a period of $1/f$, we obtain 
\begin{equation}
\begin{split}
  \frac{d\Psi}{dt}
  =&
  \gamma
  H_{\rm appl}
  -
  \pi f 
\\
  &-
  \frac{\gamma H_{\rm Ka}}{4}
  \cos\Theta
  \sin 2\Psi
  +
  \frac{\alpha\gamma H_{\rm Ka}}{4}
  \cos 2\Psi, 
  \label{eq:LLG_psi}
\end{split}
\end{equation}
where we assume that $\Theta$ is approximately constant (or $\Theta$ here might be regarded as its averaged value in the oscillation state). 
We also note that the term proportional to $\alpha H_{\rm Ka}$ is kept as the fourth term on the right-hand side of Eq. (\ref{eq:LLG_psi}), although the damping constant $\alpha$ is usually small. 
This is because another term proportional to $H_{\rm Ka}$, corresponding to the third term of Eq. (\ref{eq:LLG_psi}), has a factor $\cos\Theta=m_{x}$, which will be close to zero near the parametric oscillation state, as can be seen in Fig. \ref{fig:fig1}(b); therefore, it is not clear whether the fourth term in Eq. (\ref{eq:LLG_psi}) is sufficiently small enough to ignore compared with the third term.  
Note that we can introduce a potential $U$, satisfying $d\Psi/dt=-\partial U/\partial \Psi$, from Eq. (\ref{eq:LLG_psi}) as 
\begin{equation}
\begin{split}
  U
 & =
  -\left(
    \gamma
    H_{\rm appl}
    -
    \pi f 
  \right)
  \Psi 
  -
  \frac{\gamma H_{\rm Ka}}{8}
  \cos\Theta
  \cos 2\Psi
  -
  \frac{\alpha \gamma H_{\rm Ka}}{8}
  \sin 2 \Psi
\\
  &=
  -\left(
    \gamma
    H_{\rm appl}
    -
    \pi f 
  \right)
  \Psi 
  -
  \frac{\gamma H_{\rm Ka}}{8}
  \sqrt{\alpha^{2}+\cos^{2}\Theta}
  \cos 
  \left(
    2\Psi
    -
    \delta
  \right), 
  \label{eq:potential} 
\end{split}
\end{equation}
with $\cos\delta=\cos\Theta/\sqrt{\alpha^{2}+\cos^{2}\Theta}$ and $\sin\delta=\alpha/\sqrt{\alpha^{2}+\cos^{2}\Theta}$. 
When the frequency $f$ of the microwave voltage equals to twice the value of the Lamor frequency $f_{\rm L}=\gamma H_{\rm appl}/(2\pi)$, i.e., $f=2f_{\rm L}$, the potential $U$ becomes a double-well potential described by $-\cos(2\Psi-\delta)$. 
In this case, the potential has minima at $\Psi=\Psi_{0}=\delta/2$ and $\Psi_{0}+\pi$, and the values of the potential at these points are the same, i.e., the double-well potential is symmetric; see Fig. \ref{fig:fig4}(a), where such a symmetric potential is schematically shown by a red solid line. 
This is consistent with the results shown in Fig. \ref{fig:fig2}(b), where the two possible phases in the parametric oscillation state differ by nearly $\pi$. 
Note also that the result indicates the validity not to ignore the fourth term in Eq. (\ref{eq:LLG_psi}) mentioned above; if we neglect this term proportional to $\alpha$, $\delta$ becomes zero, i.e., $\Psi_{0}=0$ and $\pi$ are predicted to be the value of $\Psi$ in the parametric oscillation state; however, it contradicts with the numerical simulation shown as an inset of Fig. \ref{fig:fig2}(b), where the peak positions of $m_{z}$ and the microwave voltage are slightly different. 
Thus, the fourth term in Eq. (\ref{eq:LLG_psi}) should be kept. 


Equation (\ref{eq:potential}) also indicates that the potential becomes asymmetric due to the term $-(\gamma H_{\rm appl}-\pi f)\Psi$ when $f \neq 2 f_{\rm L}$, i.e., the depths of two minima are different; see Fig. \ref{fig:fig4}(a), where such an asymmetric potential is schematically shown by a blue solid line for the case of $f=2 f_{\rm L}\times 0.98 $. 
Note also that the depth of the local minima becomes shallow due to the term $-(\gamma H_{\rm appl}-\pi f)\Psi$. 
For example, while the depth of the normalized potential, $U/(\gamma H_{\rm appl})$, for $f=2f_{\rm L}$ is on the order of $10^{-4}$, that for $f \neq 2f_{\rm L}$ is nearly one order of magnitude shallower than it; see Fig. \ref{fig:fig4}(a) and its inset, as well as \ref{sec:AppendixB}. 
Remind that Eq. (\ref{eq:LLG_psi}) was obtained after applying an averaging technique to a fast variable, i.e., while, for example, $m_{x}=\cos\Theta$ is not a constant in a steady state [see Fig. \ref{fig:fig1}(b)], we replace it in Eq. (\ref{eq:LLG_psi}) with an averaged value. 
The component $m_{x}$ oscillates around the averaged value and thus, the oscillation trajectory of the magnetization around the $x$ axis is slightly distorted from a circle. 
This is because the magnetic fields in the $y$ and $z$ directions are different, i.e., the oscillating magnetic field due to the VCMA effect appears in the $z$ direction only. 
A small-amplitude oscillation of $m_{x}$, which was hidden by the averaging technique, acts as a perturbation and prevents to stabilize $\Psi$ to the minima, i.e., $\Psi$ also shows an oscillation around local minima. 
When the depth of the local minimum is shallow, this perturbation might move $\Psi$ to a deeper local minimum. 
In this case, the phase of the magnetization will be uniquely locked to a value giving a deeper local minimum of the potential.  


This point is mathematically formulated as follow. 
The steady state solutions of Eq. (\ref{eq:LLG_psi}) are generally given as $\Psi_{1}=\Psi_{0}$, $\Psi_{2}=(\pi/2)-\Psi_{0}$, $\Psi_{3}=\Psi_{1}+\pi$, and $\Psi_{4}=\Psi_{2}+\pi$, where $\Psi_{0}$ is 
\begin{equation}
  \Psi_{0}
  =
  \frac{1}{2}
  \sin^{-1}
  \left[
    \frac{4(\gamma H_{\rm appl}-\pi f)}{H_{\rm Ka} \sqrt{\cos^{2}\Theta+\alpha^{2}}}
  \right]
  +
  \frac{\delta}{2}.
\end{equation}
When $f=2f_{\rm L}$, $\Psi_{0}$ reproduces $\delta/2$ mentioned above. 
The potential $U$ has minima at $\Psi=\Psi_{1}$ and $\Psi_{3}$ and has maxima at $\Psi=\Psi_{2}$ and $\Psi_{4}$. 
Let us define the depth of the potential near $\Psi=\Psi_{1}$ ($\Psi_{3}$) as $\Delta U_{1}=U(\Psi=\Psi_{2})-U(\Psi=\Psi_{1})$ [$\Delta U_{3}=U(\Psi=\Psi_{2})-U(\Psi=\Psi_{1})$]. 
When $f\neq 2 f_{\rm L}$, these depths are different as (see also \ref{sec:AppendixB}) 
\begin{equation}
  \Delta U_{3}
  -
  \Delta U_{1}
  =
  \frac{\pi \left(\gamma H_{\rm appl}-\pi f \right)}{2}. 
  \label{eq:potential_difference}
\end{equation}
Therefore, when $f<2 f_{\rm L}=\gamma H_{\rm appl}/\pi$, the depth of the potential well near $\Psi=\Psi_{3}$ is higher than that near $\Psi=\Psi_{1}$. 
Thus, it is relatively easy to move from $\Psi=\Psi_{1}$ to $\Psi=\Psi_{3}$ by a perturbation but is difficult to return from $\Psi=\Psi_{3}$ to $\Psi=\Psi_{1}$. 
Thus, $\Psi$ will stay at $\Psi_{3}$ in this case. 
On the other hand, when $f>2f_{\rm L}$, $\Psi$ will stay at $\Psi_{1}$. 
Accordingly, when the frequency $f$ of the microwave voltage differs from $2f_{\rm L}$, the phase of the magnetization is uniquely locked. 

Starting from $f \neq 2 f_{\rm L}$, let us consider to sweep the frequency $f$ to $2 f_{\rm L}$. 
Initially, $\Psi$ is locked to a unique value, $\Psi_{1}$ or $\Psi_{3}$, depending on whether $f>2f_{\rm L}$ or $f<2f_{\rm L}$. 
When $f$ becomes $2f_{\rm L}$, the depths of the potential well, $\Delta U_{1}$ and $\Delta U_{3}$, become deeper than those for $f \neq 2 f_{\rm L}$, as mentioned above.
Therefore, $\Psi$ hardly moves between two wells after $f$ becomes $2f_{\rm L}$; rather, $\Psi$ will stay $\Psi_{1}$ or $\Psi_{3}$, which is determined by whether $f$ is initially higher or lower than $2f_{\rm L}$. 
Accordingly, the phase will be uniquely locked by using a sweeping frequency. 
Let us verify this proposal, as shown in the next subsection. 




\subsection{Results of numerical simulation}
\label{sec:Results of numerical simulation}

In this work, we use the following form of the sweeping frequency, 
\begin{equation}
  f 
  =
  \begin{cases}
    2 r f_{\rm L} & (t \le \tau_{\rm w}) \\
    2 f_{\rm L} \left[(1-r) \tanh\left(\frac{t-\tau_{\rm w}}{\tau_{\rm s}}\right) + r\right] & (t > \tau_{\rm w})
  \end{cases}, 
  \label{eq:frequency_sweeping}
\end{equation}
where $r$ is a rate of the microwave frequency with respect to $2f_{\rm L}$. 
Time $\tau_{\rm w}$ is a waiting time to keep $f$ to $2rf_{\rm L}$, while time $\tau_{\rm s}$ is a sweeping time to change $f$ to $2f_{\rm L}$. 
In Fig. \ref{fig:fig4}(b), we show an example of Eq. (\ref{eq:frequency_sweeping}) with $r=0.98$, $\tau_{\rm w}=500$ ns, and $\tau_{\rm s}=200$ ns, where the frequency of the microwave voltage is initially $2rf_{\rm L}$ and finally $2f_{\rm L}$. 
In Fig. \ref{fig:fig4}(c), we show an example of the solution $m_{z}$ when the frequency is $2rf_{\rm L}$. 
The oscillation amplitude is slightly smaller than that for $f=2f_{\rm L}$; see Fig. \ref{fig:fig2}(b), for comparison. 
We confirm that the phases of all $100$ samples in this case are the same, and even after the frequency is changed to $2f_{\rm L}$, the phases in the samples are kept identical.

In Fig. \ref{fig:fig4}(d), we summarize the dependence the locked rate on the waiting time $\tau_{\rm w}$ and the sweeping time $\tau_{\rm s}$, where their minimum values are set to be $5$ ns. 
When $\tau_{\rm w}$ ($\tau_{\rm s}$) is varied, $\tau_{\rm s}$ ($\tau_{\rm w}$) is set to be 10 ($500$) ns. 
The locked rate becomes $1$ for $\tau_{\rm w} \gtrsim 100$ ns and for $\tau_{\rm s} \ge 5$ ns. 
The results indicate that using a sweeping frequency is efficient to lock the phase of the oscillation uniquely. 
On the other hand, for $\tau_{\rm w}\lesssim 100$ ns, the locked rate is smaller than $1$, which indicates that the phase is not uniquely locked from two possible values. 
This result might relate to the fact that such a waiting time is comparable or shorter than a transition time from an initial state to a steady state. 
The transient time to a steady oscillation state is on the order of $100$ ns, as implied from Fig. \ref{fig:fig1}(b). 
When the waiting time is comparable or shorter than such a value, the frequency of the microwave voltage starts to change before the magnetization reaches a steady state. 
In this case, the phase is not uniquely locked by the microwave voltage having the frequency of $2rf_{\rm L}$; therefore, even after the frequency is changed to $2f_{\rm L}$, the phase is not uniquely locked. 
Regarding these results, the waiting time is a relatively important factor to lock the phase of the magnetization.



\subsection{Role of thermal activation}
\label{sec:Role of thermal activation}

The remaining issue is a role of thermal activation in the parametric oscillation state. 
As mentioned above, thermal activation is only included in the evaluation of the initial state. 
In reality, however, it also affects the magnetization dynamics in the presence of the microwave voltage. 
It is known that thermal activation induces a transition between two minima in a double-well potential \cite{brown63}. 
As a result, the phase cannot stay stably in one of the two minima. 
In particular, when the potential shape is symmetric, the probability to find a system in one of the two stable states will be 50 \%, i.e., the locked rate will be $0.5$. 


\begin{figure}
\centerline{\includegraphics[width=1.0\columnwidth]{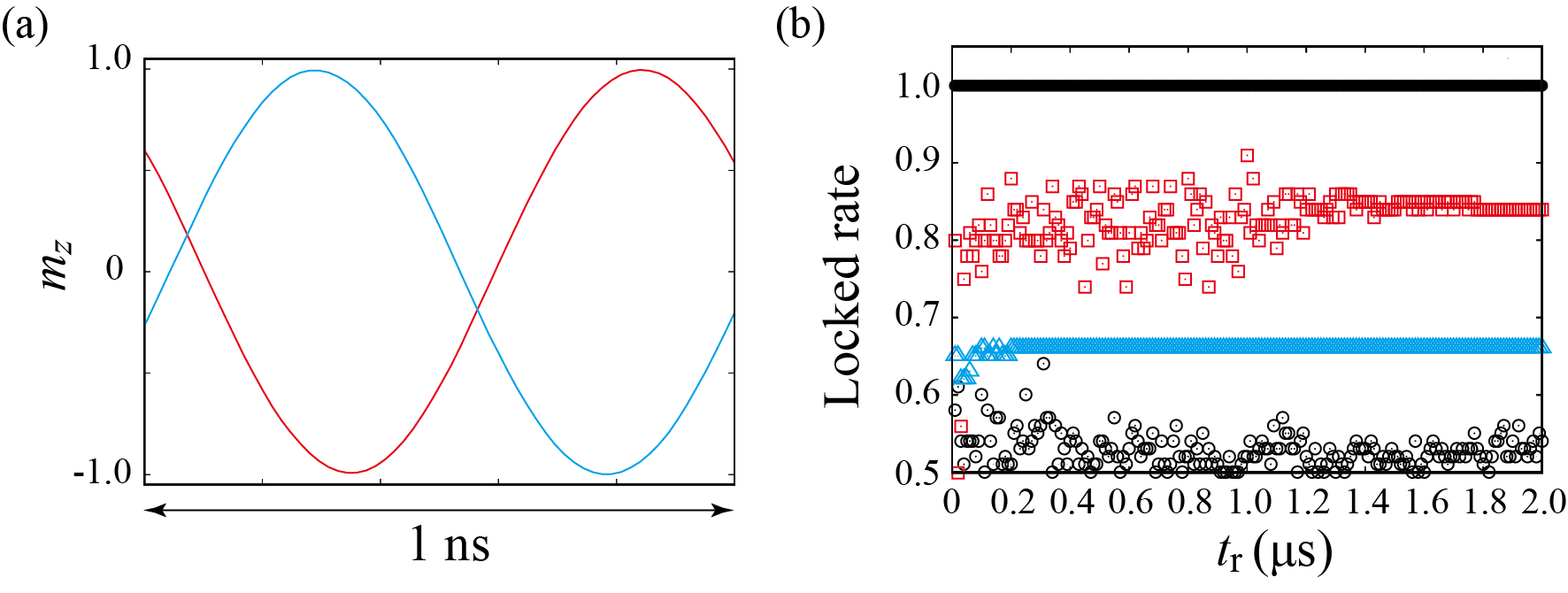}}
\caption{
           (a) Examples of oscillation of $m_{z}$, where thermal activation is included in the LLG equation even after the microwave voltage is applied. 
           (b) Locked rates as a function of running time $t_{\rm r}$ for various oscillating magnetic anisotropy field $H_{\rm Ka}$ and volume $V$. 
               The thickness $d$ of the ferromagnet is fixed to $1.1$ nm, while the radius $r$ is changed ($V=\pi r^{2}d$). 
               The values of $H_{\rm Ka}$ and $r$ are $100$ Oe and $50$ nm for black open circles, $500$ Oe and $50$ nm for blue triangles, $100$ Oe and $400$ nm for red squares, and $500$ Oe and $400$ nm for black solid circles. 
         \vspace{-3ex}}
\label{fig:fig5}
\end{figure}


We confirm it by solving the LLG equation by adding a random torque even after applying a microwave voltage, where the frequency of the voltage obeys Eq. (\ref{eq:frequency_sweeping}). 
In Fig. \ref{fig:fig5}(a), we show examples of the magnetization oscillation for $T=300$ K, $\tau_{\rm w}=200$ ns, and $\tau_{\rm s}=100$ ns. 
It shows that, even at finite temperature, the phase of the magnetization oscillation has two possible values. 
The values of the phase between two samples is slightly different from $\pi$, which might be due to an instantaneous phase disturbance by thermal activation and/or time after saturating the frequency of the microwave voltage is still too short to lock the phase. 
To evaluate the locked rate at finite temperature, we solve the LLG equation with nonzero random torque for $0 \le t \le \tau_{\rm w}+\tau_{\rm s}+t_{\rm r}$ (see also \ref{sec:AppendixA} for the definition of the locked rate at finite temperature). 
Here, $t_{\rm r}$ is a characteristic running time for applying the microwave voltage after the frequency becomes sufficiently close to $2f_{\rm L}$. 
In Fig. \ref{fig:fig5}(b), we summarize the locked rate as a function of the running time. 
It shows that the locked rate, shown by black open circles, is approximately $0.5$ for a wide range of the running time (see also \ref{sec:AppendixA}). 
Therefore, even if we use the sweeping frequency, the phase will not be uniquely locked when thermal activation exists. 

A solution to this issue might be to use a large amplitude $H_{\rm Ka}$ for the oscillating perpendicular magnetic anisotropy. 
As implied from Eq. (\ref{eq:potential}), $H_{\rm Ka}$ determines the depth of the minima of the potential. 
In fact, the depth of the potential well introduced in Sec. \ref{sec:Theoretical aspect of proposal} for $f=2f_{\rm L}$ is $\Delta U_{1}=\Delta U_{2}=\gamma H_{\rm Ka}\cos^{2}\Theta/(4 \sqrt{\alpha^{2}+\cos^{2}\Theta})$ (see also \ref{sec:AppendixB}). 
Therefore, a transition between minima may possibly be suppressed when a large $H_{\rm Ka}$ is induced by the VCMA effect. 
Note that and an enhancement of the VCMA efficiency is rapidly growing, and now the controllable range of the perpendicular magnetic anisotropy field is on the order of 1.0 kOe  \cite{maruyama09,shiota09,nozaki10,endo10,shiota11,wang11,shiota12,kanai12,shiota13,grezes16,shiota17,nozaki18,okada18}. 
Therefore, while the value $H_{\rm Ka}=100$ Oe used here is smaller than that used in an experiment \cite{yamamoto20} ($H_{\rm Ka}$ in the experiment of Ref. \cite{yamamoto20} was about $503$ Oe), we examine to evaluate the locked rate for a relatively large value, $H_{\rm Ka}=500$ Oe. 
We found that the locked rate becomes larger than $0.6$, as shown by blue triangles in Fig. \ref{fig:fig5}(b).
This is an approximately $10$ \% improvement from the results obtained for $H_{\rm Ka}=100$ Oe [shown by black open circles in Fig. \ref{fig:fig5}(b)]. 
It also indicates that the locked rate is unchanged for a long running time limit. 
It indicates that a large $H_{\rm Ka}$ contributes to fix the phase after the frequency of the microwave voltage becomes $2f_{\rm L}$. 
This is consistent with the above estimation of the depth for the potential well, where it is proportional to $H_{\rm Ka}$ when $f=2f_{\rm L}$. 
On the other hand, the locked rate depends on the running time for its short limit. 
This is because the frequency does not perfectly saturate to $2f_{\rm L}$ yet, and thus, the depth of the potential is relatively shallow, as discussed in Sec. \ref{sec:Theoretical aspect of proposal}. 

Another solution to enhance the locked rate is to suppress the effect of thermal activation by using a large-volume sample. 
For example, the volume $V$ of a device in Ref. \cite{shiota17} is relatively large due to a large radius of $400$ nm.
The red squares in Fig. \ref{fig:fig5}(b) shows the locked rate for a device with the radius of $400$ nm. 
The value of $H_{\rm Ka}=100$ Oe is the same with that used to obtain the black open circles in the same figure. 
The locked rate is found to become close to $0.9$. 
In addition, by combining two approaches, i.e., large oscillating magnetic anisotropy field $H_{\rm Ka}$ and volume, the locked rate reaches $1$, as shown by black solid circles in Fig. \ref{fig:fig5}(b). 
The results indicate that there are various approaches to enhance the locked rate. 

\section{Conclusion}
\label{sec:Conclusion}

In conclusion, we studied the parametric magnetization oscillation induced by the microwave-voltage driven VCMA effect. 
As reported in the previous work \cite{yamamoto20}, the magnetization oscillates with the Larmor frequency $f_{\rm L}$ when the frequency $f$ of the microwave voltage is $2f_{\rm L}$. 
Since the frequency of the microwave voltage is twice the value of the magnetization oscillation, it does not uniquely fix the phase of the magnetization; instead, there are two possible phases locked by the microwave voltage. 
The phenomenon was confirmed by the numerical simulation of the LLG equation with various initial conditions. 
To lock the phase uniquely, we examined two approaches. 
The first one is to enhance the perpendicular magnetic anisotropy before applying the microwave voltage. 
We found that, although this method contributes to suppress the distribution of the initial state, it is not effective enough to lock the frequency uniquely. 
A locked rate, which quantifying a uniqueness of the phase in the parametric oscillation state, is distributed even for relatively large perpendicular magnetic anisotropy fields. 
The other approach is to use a sweeping frequency, where the frequency of the microwave voltage slightly differs from $2f_{\rm L}$ in the initial state. 
Using this method, the phase is locked uniquely for $f\neq 2f_{\rm L}$ due to an asymmetric shape of a potential. 
Even after the frequency $f$ is changed to $2f_{\rm L}$, the phase is still locked. 
The usefulness of this method for phase locking was verified by the numerical simulation of the LLG equation and the quantification of the locked rate. 
Thermal activation, however, causes random transition between two phases. 
The issue will be solved by using a large oscillating magnetic anisotropy field and/or a large-volume sample.


\section*{Data availability}

Data will be made available on request. 

\section*{Acknowledgement} 

The author is grateful to Takayuki Nozaki for discussion. 
The work is supported by JSPS KAKENHI Grant Number 20H05655.


\appendix


\section{Details of numerical simulation}
\label{sec:AppendixA}

In this work, the LLG equation is solved by the $4^{\rm th}$ order Runge-Kutta method with a time increment $\Delta t =1$ ps. 

The locked rates in Secs. \ref{sec:Suppression of initial distribution} and \ref{sec:Results of numerical simulation} are estimated as follows. 
Let us denote the solution of the LLG equation for the $\ell$th initial condition ($\ell=1,2,\cdots,100$) as $\mathbf{m}_{\ell}(t)$. 
In particular, the solution $\mathbf{m}_{1}$ for the $1$st sample is used as a reference sample. 
We solve the LLG equation for $0 \le t \le t_{\rm max}$ and evaluate the difference between $\mathbf{m}_{\ell}$ ($\ell=2,3,\cdots,100$) and the reference sample $\delta_{\ell}=|\mathbf{m}_{\ell}(t_{\rm max})-\mathbf{m}_{1}(t_{\rm max})|$, where $t_{\rm max}$ is $5.0$ $\mu$s in this work. 
Since there are only two possible values of the phase, we can conclude that the $\ell$th sample has the same phase with the reference one when $\delta_{\ell}$ is sufficiently small. 
On the other hand, if $\delta_{\ell}$ is large, it means that two samples have different phases. 
Although we use a condition $\delta_{\ell}\le \epsilon$ with $\epsilon=10^{-5}$ for the judgement of the definition as the same phase, the value of $\epsilon$ depends on that of $t_{\rm max}$. 
This is because $\delta_{\ell}$ becomes small as $t_{\rm max}$ increases when two samples have the same phase; therefore, when $t_{\rm max}$ is shorter than $5.0$ $\mu$s, for example, $\epsilon=10^{-5}$ might be too small to judge as the same phase automatically. 
Because of the oscillation amplitude of $m_{z}$ in the parametric oscillation state is close to $1$, as shown in Fig. \ref{fig:fig2}(b), it might be enough to use $\epsilon=1$ for the judgement of the same phase. 
Let us denote the number of the samples satisfying $\delta_{\ell}\le \epsilon$ as $n$. 
If $n\ge 50$, more than a half of the $100$ samples have the same phase with the reference sample. 
In this case, the locked rate is defined as $(n+1)/100$, where $+1$ indicates the reference sample. 
On the other hand, if $n<50$, $100-(n+1)$ samples have the opposite phase with respect to the reference sample. 
In this case, the locked rate is defined as $1-[(n+1)/100]$. 

The locked rate in Sec. \ref{sec:Role of thermal activation} is evaluated similarly. 
In this case, one might consider that $t_{\rm max}$ should be $\tau_{\rm w}+\tau_{\rm s}+t_{\rm r}$. 
It should, however, be noted that the random torque might result in a large instantaneous $\delta_{\ell}$ even if the $\ell$th sample has the same phase with the reference sample. 
In such a case, the judgement whether the sample has the same phase or not with respect to the reference sample might become ambiguous. 
Therefore, after solving the LLG equation with the random torque for $0 \le t \le \tau_{\rm w}+\tau_{\rm s}+t_{\rm r}$, we solve it continuously for $2.0$ $\mu$s by removing the random torque. 
Then, we evaluate $\delta_{\ell}$ and judge whether the $\ell$th sample has the same or opposite phase with respect to the reference sample. 
In this procedure, we assume that the phase at $t=\tau_{\rm w}+\tau_{\rm s}+t_{\rm r}$ is unchanged after thermal activation is removed. 


\section{Depth of potential}
\label{sec:AppendixB} 

To derive Eq. (\ref{eq:potential_difference}), we estimate the values of the potential at the four steady-state points as 
\begin{equation}
\begin{split}
  U(\Psi=\Psi_{1})
  =&
  -\left(
    \gamma H_{\rm appl}
    -
    \pi f 
  \right)
  \Psi_{0}
  -
  \frac{\gamma H_{\rm Ka}}{8}
  \sqrt{\alpha^{2}+\cos^{2}\Theta}
  \sqrt{
    1
    -
    \frac{16 (\gamma H_{\rm appl}-\pi f)^{2}}{\gamma^{2}H_{\rm Ka}^{2} (\alpha^{2}+\cos^{2}\Theta)}
  },
\end{split}
\end{equation}
\begin{equation}
\begin{split}
  U(\Psi=\Psi_{2})
  =&
  -\frac{(\gamma H_{\rm appl}-\pi f)}{2}
  \left\{
    \cos^{-1}
    \left[
      \frac{4 (\gamma H_{\rm appl}-\pi f)}{\gamma H_{\rm Ka} \sqrt{\alpha^{2}+\cos^{2}\Theta}}
    \right]
    +
    \cos^{-1}
    \left[
      \frac{\alpha}{\sqrt{\alpha^{2}+\cos^{2}\Theta}}
    \right]
  \right\}
\\
  &
  +
  \frac{\gamma H_{\rm Ka}}{8}
  \sqrt{\alpha^{2}+\cos^{2}\Theta}
  \cos
  \left\{
    \sin^{-1}
    \left[
      \frac{4 (\gamma H_{\rm appl}-\pi f)}{\gamma H_{\rm Ka} \sqrt{\alpha^{2}+\cos^{2}\Theta}}
    \right]
    +
    \sin^{-1}
    \left[
      \frac{\alpha}{\sqrt{\alpha^{2}+\cos^{2}\Theta}}
    \right]
  \right\}
\end{split}
\end{equation}
\begin{equation}
\begin{split}
  U(\Psi=\Psi_{3})
  =&
  U(\Psi=\Psi_{1})
  -
  \left(
    \gamma H_{\rm appl}
    -
    \pi f
  \right)
  \pi, 
\end{split}
\end{equation}
\begin{equation}
\begin{split}
  U(\Psi=\Psi_{3})
  =&
  U(\Psi=\Psi_{2})
  -
  \left(
    \gamma H_{\rm appl}
    -
    \pi f
  \right)
  \pi, 
\end{split}
\end{equation}
From these values, $\Delta U_{1}=U(\Psi=\Psi_{2})-U(\Psi=\Psi_{1})$ and $\Delta U_{3}=U(\Psi=\Psi_{2})-U(\Psi=\Psi_{3})$, as well as $\Delta U_{3}-\Delta U_{1}$, are evaluated. 
Since the mathematical expressions of $\Delta U_{1}$ and $\Delta U_{3}$ are complex, we estimate the depth numerically. 
Using the values of the parameters in the main text and the figure caption of Fig. \ref{fig:fig4}(a), we find that $\Delta U_{1}/(\gamma H_{\rm appl})\simeq 8.7\times 10^{-5} \simeq 10^{-4}$ for $f=2f_{\rm L}$ and $\Delta U_{1}/(\gamma H_{\rm appl})\simeq 1.5\times 10^{-5}$ for $f = 2 f_{\rm L}\times 0.98$.

\end{document}